\documentclass[conference]{IEEEtran}

\usepackage{cite}
\usepackage[pdftex]{color}
\usepackage{graphicx,url}
\usepackage{graphicx}
\usepackage[utf8]{inputenc}  
\usepackage{multirow}

\usepackage{amsmath}
\usepackage[T1]{fontenc}
\usepackage{bm}
\ifCLASSINFOpdf
\else
\fi
\begin{document}

\title{Cognitive Management of Bandwidth Allocation Models with Case-Based Reasoning - Evidences Towards Dynamic BAM Reconfiguration}

\author{\IEEEauthorblockN{Eliseu M. Oliveira}
\IEEEauthorblockA{
Salvador University - UNIFACS \\ 
Email: eliseu@gmail.com}
\and
\IEEEauthorblockN{Rafael Freitas Reale}
\IEEEauthorblockA{
Instituto Federal da Bahia - IFBA\\ 
Email: reale@ifba.edu.br}
\and
\IEEEauthorblockN{Joberto S. B. Martins}
\IEEEauthorblockA{
Salvador University - UNIFACS\\ 
Email: joberto.martins@gmail.com}}

\maketitle

\begin{abstract}
  Management is a complex task in today's heterogeneous and large scale networks like Cloud, IoT, vehicular and MPLS networks. Likewise, researchers and developers envision the use of artificial intelligence techniques to create cognitive and autonomic management tools that aim better assist and enhance the management process cycle. Bandwidth allocation models (BAMs) are a resource allocation solution for networks that need to share and optimize limited resources like bandwidth, fiber or optical slots in a flexible and dynamic way. This paper proposes and evaluates the use of Case-Based Reasoning (CBR) for the cognitive management of BAM reconfiguration in MPLS networks. The results suggest that CBR learns about bandwidth request profiles (LSPs requests) associated with the current network state and is able to dynamically define or assist in BAM reconfiguration. The BAM reconfiguration approach adopted is based on switching among available BAM implementations (MAM, RDM and ATCS). The cognitive management proposed allows BAMs self-configuration and results in optimizing the utilization of network resources.
\end{abstract}


%\IEEEpeerreviewmaketitle

\section{Introduction and Motivation}

The current scenario of communication networks, such as 5G, Cloud, IoT, vehicular and MPLS networks, is marked by a wide variety and large distribution of services, applications and users alongside with a high volume of data exchanges with heterogeneous quality assurance requirements (SLA, QoE, QoS) \cite{elsawy_virtualized_2015}. 

These actual networks are highly heterogeneous,  have to deal with an exponential growth in the number of users, have a huge amount of data to process and extract management knowledge, are highly dynamic in terms of user's demands and are subject to failure \cite{mahmoud_cognitive_2007}.

It is also a fact that the research and developing communities have struggled in the last years to provide adaptable management solutions towards autonomic and self-managing networks. Autonomic solutions aim to reduce human intervention in complex management tasks and engineer accurate knowledge, possibly on-the-fly, to better support the management process  \cite{bezerra_network_2014}.

In this new and ever increasing complex context, management and autonomic management do require a blueprint. Aligned with this perspective, one of the actual trends in network management is to explore the concept of cognitive management in which artificial intelligence techniques are used to process management data, extract knowledge and infer decisions \cite{rendon_machine_2018}.

Resource allocation for communications has been a challenging management task for decades \cite{hui_resource_1988}. Current resource allocation proposals do reflect the aforementioned communications network scenario evolution and inherent requirements. In effect, resource allocation methods, eventually embedded in a more general autonomic management solution, must support heterogeneous and large scale networks, must have dynamic and adaptable capabilities, should preferably guarantee on-the-fly computation and, eventually, should be distributed \cite{song_game-theoretic_2014} \cite{lagunas_resource_2015} \cite{singh_qos-aware_2015} \cite{cordeschi_reliable_2015} \cite{manvi_resource_2014} \cite{zhang_resource_2015} \cite{xiao_simultaneous_2004}.

Bandwidth allocation models (BAMs) are a solution for resource allocation.  In summary, BAMs allow the definition of application or traffic classes and control the distribution and sharing of resources among them  \cite{bezerra_network_2014}. BAMs can effectively optimize resource allocation for the target networks by either reconfiguring its operational parameters or switching among BAM distinct models (MAM, RDM and ATCS) \cite{reale_preliminary_2014}.

BAM reconfiguration and BAM switching are a challenging management task that must be dynamically orchestrated considering network policies, current network traffic demand and a huge volume of management state information \cite{reale_preliminary_2014}.

We propose in this paper a cognitive management approach (BAMCBR) based on Case-Based Reasoning (CBR) for the dynamic reconfiguration (switching) of bandwidth allocation models in MPLS networks.

The motivation is to develop a CBR-based autonomic management solution for BAMs that reduces human intervention in the management process and allow the optimization of resources (bandwidth) in MPLS networks.

This article is organized as follows: section 2 discusses the related work and section 3 details the cognitive management approach with CBR. The BAMCBR module implementation is discussed in section 4 and section 5 presents its proof of concept. Finally, section 6 indicates the final considerations.

\section{Related work}

Resource allocation management for communications has been researched for many years \cite{hui_resource_1988}. Despite years of struggle, investment and results achieved, the subject keeps its importance mainly due to the communication network's evolution.

Aligned with new communication network (5G, Cloud, IoT, vehicular, MPLS, SDN, other) requirements, actual resource allocation management research considers multiobjective optimization \cite{xiao_simultaneous_2004}, envisions scalability for large and highly distributed networks \cite{guizani_large-scale_2015}, explore distributed and adaptable capabilities \cite{cordeschi_reliable_2015} and considers computational efficiency and dynamics towards on-the-fly management solutions \cite{}.

As discussed in \cite{rendon_machine_2018} and \cite{mahmoud_cognitive_2007}, cognitive management using artificial intelligent techniques became more recently a relevant asset for communications management. Cognitive management has been applied using different AI techniques for cellular systems \cite{guizani_large-scale_2015}, vehicular networks \cite{cordeschi_reliable_2015}, satellite communications \cite{lagunas_resource_2015}, small cells mobile systems \cite{zhang_resource_2015} and, more extensively, to cognitive radio \cite{wu_spectrum_2010}.

Bandwidth allocation model research has been focused on developing new models that, somehow, complement the basic ones: i) MAM proposed by \cite{faucher_maximum_2005}, ii) RDM discussed in  \cite{da_costa_pinto_neto_rdm-like_2008}; and iii) the AllocTCSharing (ATCS) model proposed by \cite{reale_alloctc-sharing:_2011}. The research group associated with this work has proposed 2 new basic BAM: i) the AllocTCSharing (ATCS) model \cite{reale_alloctc-sharing:_2011}; and ii) the Generalized Bandwidth Allocation Model (GBAM) \cite{reale_g-bam:_2014}. Some additional hybrid models that fundamentally provide some extra capabilities to existing models have also been proposed by \cite{pistek_-mar:_2015} \cite{tata_cam:_2013} and  \cite{da_costa_p._neto_adapt-rdm_2008}.

To the best of our knowledge, none of these new resource allocation management approaches has takled yet the problem of cognitive management applied for BAM-based resource allocation.

The proposed BAMCBR module is part of the REPAF project \cite{martins_repaf_2017} in which BAMs and cognitive management are investigated to allocate resources in MPLS, IoT \cite{moraes_publish/subscribe_2018}, EON (Elastic Optical Networks) \cite{reale_evaluating_2017} and NFV (Network Function Virtualization) networks.

\section{Cognitive Management of BAMs with Case-Based Reasoning - Basic Approach and Issues}

Cognitive management principle applied to autonomic management, as discussed in \cite{rendon_machine_2018}, state that "management \textit{actions} should be \textit{learned} from the network environment, \textit{reasoned} and eventually \textit{adapted} while respecting management goals and requirements".

The BAM cognitive management approach with CBR is depicted in Figure 1. There are 3 components that interact to learn and actuate: the BAM module, the BAMCBR module and the target MPLS network. The BAM module acts as a broker for MPLS users requesting LSPs setup and allocates the required bandwidth (network links resource) for LSP setup. The BAMCBR module learns from the MPLS network and infers the necessary BAM reconfiguration actions.

\begin{figure}[ht]
\centering
\includegraphics[width=0.4\textwidth]{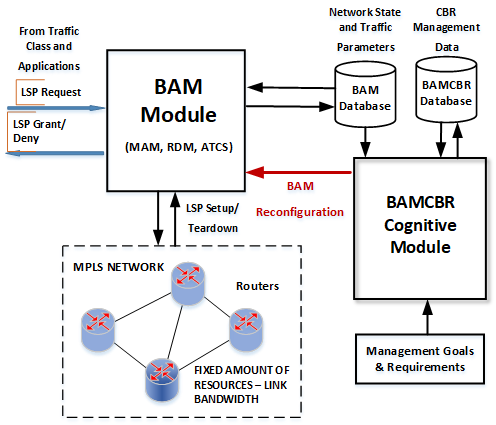}
\caption{BAM Cognitive Management Approach}
\end{figure}

Considering the approach depicted in Figure 1, the following issues to discuss are: i) what aspects of BAM operation are configurable and; ii) how adequate CBR technique is for BAM reconfiguration.

\subsection{BAM Configuration Management Issues}

Firstly, it is worth to mention that BAM is a resource allocation solution suitable for controlled networks or systems that have a fixed amount of resources and envisage to distribute and share them among users.

BAM configuration involves 3 phases: i) the definition of classes of services (TCs - Traffic Classes) with common requirements (QoS, SLA or other user/ application parameter); ii) the definition of the amount of bandwidth per class (BC - Bandwidth Constraint); and iii) BAM model configuration with an inherent behavior for resource sharing among TCs \cite{martins_uma_2015}. Traffic classes are typically static since they put together applications/ users with common network requirements. Consequently, cognitive management may effective explore two configuration alternatives: i) the amount of bandwidth defined per class; or ii) the sharing strategy among TCs. BAMCBR proposal explores the reconfiguration of the sharing strategy.

Next issue is, how resource sharing will  be explored by BAMCBR cognitive management? The answer is: changing from no sharing at all between TCs, to sharing among priority and non-priority TCs and, finally, to sharing between all TCs, independently of their priority. The switch among BAM models, as described in the following paragraphs, clarifies this approach.

There are 03 basic BAM models: (i) Maximum Allocation Model (MAM) \cite{faucher_maximum_2005}, (ii) Russian Dolls Model (RDM) \cite{da_costa_pinto_neto_rdm-like_2008} and (iii) AllocTC-Sharing (ATCS) \cite{reale_alloctc-sharing:_2011}. There are also several hybrid BAM versions like described in \cite{pistek_-mar:_2015} \cite{da_costa_p._neto_adapt-rdm_2008} and \cite{tata_cam:_2013} that will not be considered in BAMCBR proposal.

In summary, MAM model allocates bandwidth without any resource sharing between traffic classes (TCs). The RDM model allows temporary sharing of unused higher priority classes bandwidth by  lower priority traffic classes (High-to-Low - HTL strategy). The ATCS model allows the generalization of bandwidth sharing between all priority and non-priority traffic classes (LTH and HTL strategies - Low-to-High) \cite{martins_uma_2015}.

All basic and hybrid BAMs are implemented by BAMCBR using the Generalized Bandwidth Allocation Model (GBAM) that allows the configuration of all possible models, BCs and behaviors \cite{reale_g-bam:_2014}.

\subsection{CBR Evaluation and Pertinence Issues for BAM-based Cognitive Resource Allocation}

As discussed in \cite{gu_evaluating_2006}, cognitive management research is an empirical process in which a management task is selected to incorporate 
intelligence features, a management system is built incorporating these features and the system is evaluated with different management task scenarios. Cognitive systems evaluation methods include statistical evaluation, theoretical analysis, tuning evaluation, limitation evaluation and characteristic analysis, among other alternatives \cite{cohen_toward_1989}.

In the perspective of BAMCBR evaluation, \cite{gu_evaluating_2006} also points out that typical complexity of CBR application contexts makes CBR solution evaluation difficult, if not impossible.

Aligned with the aforementioned technical aspects, a BAMCBR proof of concept is the preliminary evaluation method adopted (section V), associated with an CBR pertinence characteristic analysis.

In relation to the later mentioned evaluation aspect, the most relevant CBR pertinence characteristics include: i) the fact that CBR is based on the reuse of previous problem solution "cases" to solve new ones; and ii) CBR supports adequately knowledge intensive applications. By intuition, is common sense that problems tend to recur, so new problems are often similar to previous ones.

\section{BAMCBR Implementation - Problem Determination, Learning and Reasoning Method}

BAMCBR implementation aspects like problem determination, knowledge control mapping and learning and reasoning method adopted are described in this section.

\subsection{Problem Determination and Knowledge Control}

BAMCBR uses Case-Based Reasoning (CBR) as the learning and reasoning techniques. For CBR operation, as proposed by \cite{agoulmine_autonomic_2011} \cite{dobson_survey_2006} and \cite{jennings_towards_2007}, it is necessary clearly determine the problem and how the acquired knowledge will infer actions promoting network adaptability (BAM reconfiguration).

The problem determination and knowledge control are mapped to CBR context aiming to optimize the competition of resources arbitrated by BAMs.

The problem determination consider the following aspects: i) defines context and reason for it; ii) represents the captured information or infers it; iii) specifies how to infer new knowledge from existing one to detect symptoms and make a decision; iv) allows the system to learn about new states and improve its capabilities; and v) defines a uniform approach to represent problems and associated potential solutions.

%Problem Determination: 
%\begin{itemize}
%   \item Defines context and reason for it;
%    \item Represents the captured information or infers it;
%    \item Specifies how to infer new knowledge from existing one to detect symptoms and make a decision; 
%    \item Allows the system to learn about new states and improve its capabilities; and
%    \item Defines a uniform approach to represent problems and associated potential solutions.
%\end{itemize}

The knowledge control: i) aims to determine if changes need to be made, or not, in the elements managed through policies; and ii) provides a uniform and neutral way to define control policies to govern the autonomous network decision process.

%Knowledge Control: 
%\begin{itemize}
%    \item Aims to determine if changes need to be made, or not, in the elements managed through policies; and
%    \item Provides a uniform and neutral way to define control policies to govern the autonomous network decision process.
%\end{itemize}

\subsection{BAMCBR Learning and Reasoning Method}

The BAMCBR module is based on the CBR 4R cycle defined by \cite{aamodt_case-based_1994} that encompasses a cycle of continuous reasoning composed of four main stages (Figure 2):

%\begin{figure}[ht]
%\centering
%\includegraphics[width=0.5\textwidth]{IEEEtran/imagens/module_BAMCBR.png}
%\caption{BAMCBR Module 4R Cycle}
%\label{fig:bamCBRResumo}
%\end{figure}

\begin{itemize}
    \item "Evaluation and Proposal" (Recovery) - Responsible for interpreting the data obtained and searching the knowledge base for the best alternative (configuration) for the current state of the network.
    \item "Adaptation and Use" (Reuse) - Responsible for adapting the proposed solution and applying it to the network.
    \item "Test and Review" (Revision) - Responsible for verifying the effectiveness of the solution applied and proposing changes, if necessary.
    \item "Storage and Learning" (Retention) - Responsible for assimilating the knowledge acquired in the process of analysis and planning for a fast execution in a future occurrence.
\end{itemize}

The following subsections present with greater detail the BAMCBR module operation.

\subsection{The BAMCBR 4R Cycle}

The BAMCBR cycle starts with the "Evaluation and Proposal" stage. This task can be activated in two different ways: i) reactively; or ii) proactively. 

Reactive Mode (Figure 2 - A1) - Occurs when an external entity requests an analysis and solution for the current network state, with the purpose of obtaining an improvement and/or optimization of this one. 

Proactive Mode (Figure 2 - A2) - Occurs from time to time according to the internal timer setting, to proactively check network state and, if necessary, to propose improvement actions.

%Regardless of how BAMCBR cycle is started, the "Evaluation and Proposal" task will always be performed as the first step of the cycle.

\begin{figure}[ht]
\centering
\includegraphics[width=0.4\textwidth]{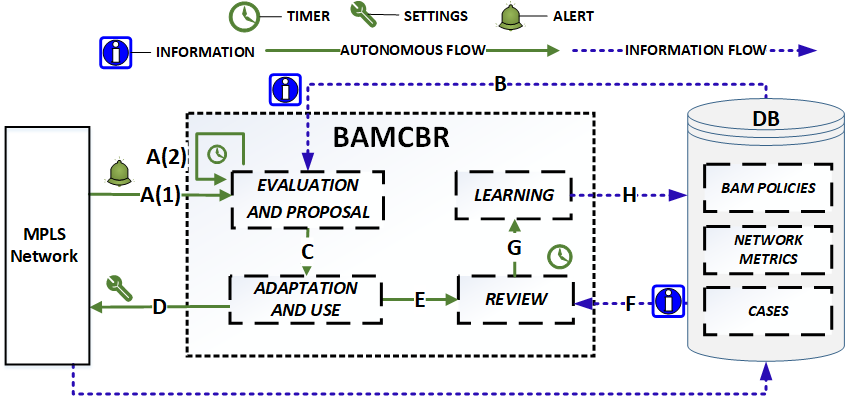}
\caption{BAMCBR Cycle}
\end{figure}

At the beginning of the cycle, after receiving an alert from the external module (Figure 2 - A1), the BAMCBR module requests from the network all values of the attributes relevant to it (preemptions, delays, LSPs, etc.) containing the characteristics that describe the current state of the network. This set of information is called "Current Problem".

With the "Current Problem", the "Evaluation and Proposal" stage checks the network status and the BAM policies defined by the manager (Figure 2 - B). This is accomplished to figure out if the current network state conforms to the manager's specifications. If the network state is not compliant, the "Evaluation and Proposal" stage is in charge to find and propose a solution to the "current problem".

Initially, "Evaluation and Proposal" stage searches in the "cases" database another previously stored network state (case) with similar characteristics to the current problem. A similarity function is used to calculate "cases" similarity \cite{m_oliveira_evaluating_2017}. The nearest cases are then returned and forwarded to the next stage as a possible solution to the current problem (Figure 2 - C).

However, if no "case" is returned from the database, the "Evaluation and Proposal" stage itself will suggest a solution. This will be accomplished taking into account the manager's previously established criteria. In an initial implementation the "Evaluation and Proposal" stage presents an arbitrary solution that is attributed to the current problem even without a correspondence in the "case" database.

Once a similar "case" is found on the "positive case" database, the "Adaptation and Utilization" stage is initiated. It receives from the first stage a tuple containing the description of the current problem (the current state of the network) and the solution to the current problem (Figure 2 - C). Then an adaptation of the solution is executed with the case "current problem". The resulting compilation is called "New Case".

Immediately after the adaptation step, the "New Case" is evaluated to verify that it meets the manager's requirements. This is done by making a comparison with the base of rejected cases. If the case has been rejected previously or marked with any manager's observation, it is considered invalid.

Finally, if the case is valid, the "Adaptation and Utilization" stage sends to the network the existing configuration in the solution field of the "New Case", as a solution to the initial alert, (Figure 2 - D). This corresponds to modify BAM's behavior (model) to meet manager's specifications. If the case is not valid then it is discarded and the whole cycle is started again.

The third stage of the cycle corresponds to "Test and Review". This stage is responsible for evaluating the efficiency of the proposed solution. It receives the "new case" containing the problem and the solution previously applied to the network. However, the evaluation does not happen immediately after the "new case" arrives at the "Test and Review" stage (Figure 2 - E). In effect, it is necessary to wait some time (timer) until the network statistics are updated with new network performance data acquired from the reconfigured network (Figure 2 - D).

Timer overflow triggers the "Testing and Revision" stage that receives new network statistics  (Figure 2 - F). With this information, a comparison is made between current network metrics and previous ones. If the network, based on the manager's criteria, presents improvement in its performance, the "new case" receives the positive status. If there were no improvements or, alternatively, if the network worsened its performance, the new case receives a negative status. In both situations the "case" is directed to the "Learning Stage".

After receiving the "new case" (Figure 2 - G), the "Learning Stage" analyzes the status of the case to know if the solution was appropriate or not for solving the problem. Regardless of the response, learning always occurs because: i) if the case was successful - it is stored in the "positive cases database" for later use; ii) if the case was unsuccessful - it is stored on the "negative cases database" so as not to be used in future occurrences.

\subsection{BAMCBR Cycle Additional Considerations}

CBR 4R cycle additional considerations are presented next to allow a deeper understanding of the learning process.

\begin{itemize}
    \item "Evaluation and Proposal" stage: if no "case" is found in the "cases database", an arbitrary solution is proposed by this stage. That is so to always guarantee a solution offer to the current problem. If this "randomly" proposed solution is evaluated as satisfactory by the 4R cycle, the case is stored on the positive cases database.  If the "randomly" proposed solution is not satisfactory, a new solution is proposed and the previous alternative is stored on the negative cases database. This process is repeated several times until a suitable solution is found.
    \item When the "Evaluation and Proposal" stage proposes a solution, the whole cycle should be executed. However, this does not need to be done in all implementations. As an option, the manager can propose a solution.
%\textcolor{red}{???????}
    \item "Testing and Review" stage: In the comparison that occurs between the current network performance metrics and the previous ones (Figure 2 - F), it is important to take into account the network profile changes (number of established LSPs, allocated bandwidth per network and per link, network current traffic demand, other). In effect, it will not be wise to evaluate the efficiency of the reconfiguration process in case the network profile has substantially changed. This could lead to an erroneous case-based learning, since it would not be possible to identify whether the network improved because the new solution was adequate or because the traffic changed.
    
%In BAMCBR implementation we ... \textcolor{red}{?????}

\end{itemize}

\section{BAMCBR Proof of Concept} \label{sec:firstpage}

A proof of concept was carried out with BAMCBR module  to verify the following capabilities:

\begin{itemize}
     \item Does CBR, as a cognitive technique, effectively learns about the MPLS network and proposes reconfiguration solutions?

    \item Even without having any previous knowledge of the MPLS network, can BAMCBR learn and propose solutions that result in network optimization?

%    \item Is BAMCBR able to identify changes in traffic profile and react, suggesting a suitable reconfiguration for the new MPLS scenario?
    
\end{itemize}

\subsection{Proof of Concept - Test Scenario Definitions}

The test scenario for BAMCBR proof of concept: i) uses the BAMSIM simulator \cite{da_costa_pinto_neto_rdm-like_2008} do simulate BAM's module operation; ii) defines the CBR domain and knowledge representation; and iii) define BAM policies.

\subsection{BAMSIM and BAM Module Simulation}   

The BAMSIM (Bandwidth Allocation Model Simulator) is a specialized simulator that simulates the BAM Module (Figure 1). BAMSIM supports all the necessary mechanisms to simulate a MPLS/DS-TE network. Functionality available include path selection algorithms and the generalized GBAM model that has the capability do be reconfigured to implement MAM, RDM and ATCS models.

\subsection{CBR Module}   
The BAMCBR cognitive module implements CBR (Figure 1). The jCOLIBRI framework was used to built the CBR implementation and integration with BAMSIM simulator. jCOLIBRI is an object-oriented framework developed in Java for the purpose of building CBR-oriented systems \cite{recio-garcia_j_2014}.

\subsection{CBR Domain Definition}

BAMCBR requires two data input sets: network state information (measurements, statistics, etc.) included in the domain definition and the policies/goals of the network manager.
%(Figure \ref{fig:BAMCBRApproach}).

The first step in the process of using CBR in any area of knowledge is the definition of the "domain" where the problem is located. Domain definition requires the definition of attributes that will be used as indexes to represent domain's knowledge. These indexes are important to the success of the CBR implementation, since it is through these indices that CBR performs the similarity search on the "cases database".

Four components were defined to represent the domain and for the calculation of similarity \cite{m_oliveira_evaluating_2017}: 

\begin{itemize}
    \item Contextual Information - Contextual attributes are related to the current network configuration and the policies defined by the network manager. They identify the context in which similarity should act and their information is acquired before the network optimization process starts. Examples of contextual attributes are: BAM currently used, network manager tolerance for network problems (preemption, devolution, blocking, etc), bandwidth defined for each BC, among others.
    \item     Measurements - These are information obtained from the network and provide a snapshot of the current state of the network. These measurements are essential to portray changes in the network profile. Some examples of measurements are: total or by TCs network utilization, total or by TCs preemption, total or by TC devolution and total or by TCs LSP blocking.
    \item   Similarity Function - It is the method used to execute the case's search in the database. One or more similarity functions can be used. The similarity function is responsible for the comparison of cases. Examples of similarity functions are: linear, ladder and nearest neighbor \cite{m_oliveira_evaluating_2017}.
    \item  BAM-CBR Problem (symptom) - The BAM-CBR problem is the symptom/alert that suggests and characterizes the current problem. These symptoms may be from emergency alerts to setups for periodic diagnoses.
\end{itemize}
The definition of the problem and associated policies is depicted in the following subsection.

\subsection{BAM Problems and Policies}
   
The BAM policy is the set of criteria and rules specified by the network manager for each network assisted "condition". 

Actions are associated with "conditions" and responses are executed when these conditions are satisfied. In the BAMCBR domain the BAM policy is the BAM-CBR problem itself. For example, in an network where the network manager defines in the policy that the high devolution number is not a relevant condition, no problem is generated. However, if in this network the number of preemption is a BAM policy action factor, an action must be taken and a problem is mapped and sent to BAMCBR.

Just like the cases in CBR, the BAM policies are also defined as a tuple Problem/ Solution. In this work the following policies were defined for the representation of BAM problems.

\begin{itemize}
    \item Problem: Network uses MAM; link utilization is low.
    \item [-] Solution - Reconfigure the network to use ATCS model.
    \item Problem: Network uses RDM; link utilization level is low; blocking rate is high.
    \item [-] Solution - Reconfigure the network to use ATCS model.
    \item Problem: Network uses RDM; link utilization level is high; preemption level is high.
    \item [-] Solution - Reconfigure the network to use MAM model.
    \item Problem: Network uses ATCS; link utilization level is high; preemption level is low; devolution level is high.
    \item [-] Solution - Reconfigure the network to use RDM model.
    \item Problem: Network uses ATCS; link utilization level is high; preemption level is high.
    \item [-] Solution - Reconfigure the network to use MAM model.
\end{itemize}

%Figure \ref{fig:ProblemaBAM-CBR} illustrates all policies with their respective actions.

%\begin{figure}[ht]
%\centering
%\includegraphics[width=0.5\textwidth]{IEEEtran/imagens/Politicas_BAM_ENG.png}
%\caption{BAMCBR Problems and Policies}
%\label{fig:ProblemaBAM-CBR}
%\end{figure}
 
%Das políticas apresentadas, para esta prova de conceitos determinamos como critérios de performance e otimização : (1) minimizar as preempções e devoluções; (2) contudo maximizar a utilização do enlace. Ou seja a taxa de preempção e devolução deverá ser próxima a taxa de preempção e devolução do MAM (zero – 0), e a taxa de utilização próxima as taxas do ATCS . Sendo assim um gestor que implemente essa política espera aproveitar o melhor do MAM e do ATCS.
 
\subsection{Network Test Topology and Input Traffic Patterns}
    The proof of concept defined is focused on  verifying, firstly, if CBR effectively learns and proposes a new reconfiguration and, secondly, if it does achieve any network optimization.
    
    At this point it is also worth to remember that BAM models control bandwidth allocation on a per-link basis (independent BAM control and allocation for each link). As such, a single link topology (point-to-point link between routers) is used to implement the target MPLS circuit. In a real network there will exist many links and the BAMCBR operation will execute independently for all links. As such, the preliminary conclusions obtained about learning capability for this simplified topology will hold true for a more complex topology on a per-link analysis.
    
    Of course, the overall performance achieved for a multiple link topology using BAMCBR can not be inferred from the results obtained with this proof of concept. This aspect is being object of further research.

    Additional network test scenario definitions are: i) three classes of traffic (TC0 - BC0 = 400M; TC1 - BC1 = 350M and TC2 - BC2 = 250M; link = 1G) for accommodate users.

%\begin{figure}[ht]
%    \centering
%    \includegraphics{IEEEtran/imagens/TabelaBCs.png}
%    \caption{BC's Configured Bandwidth}
%    \label{fig:largura_banda_BCs}
%\end{figure}
        
A set of patterns with random input traffic within each of them are generated in the simulation (Table I). Six input traffic configurations (1 hour duration each) are used. The first three input traffic patterns configurations are intended to allow bandwidth sharing between TCs and allow random cognitive management actions by BAMCBR. The last three were created to overload the network and investigate how BAMCBR behaves with these input traffic conditions. These patterns are repeated for 4 times (24h simulation). The results obtained are discussed in the following session.

The set of simulation patterns was designed incorporating previous manager's knowledge about what would be the best possible BAM model for each specific traffic pattern. This allow the verification if BAMCBR module does effectively learns about the current network status and acts reconfiguring BAMs. The last line of Table I indicates the best BAM model to be configured according manager's knowledge, expertise and perception for each traffic condition.

%Taking into account the three classes of traffic used in this simulation (TC0, TC1 and TC2) and remembering that the lower traffic classes have lower priority in relation to the larger classes (TC0 <TC1 <TC2), traffic profile 1 was constructed from low  TC1 and TC2 flow and a high TC0 flow. Notoriously traffics with this profile have better performance in networks if used from BAM RDM or ATCS. It is possible to observe this profile in the figure \ref{fig:perfil_trafego1-3} in times 0:0h to 1:0h.

%The traffic profile 2 inverts between the TC0 and TC2 with a slight increase in TC1 traffic.

%It is possible to observe that the TC2 (which only has 25\% bandwidth of the link) is using more than 50\% of the bandwidth of the link or is doing loan of TCs 0 and 1. This typical behavior is only possible using ATCS, best suited for this profile. This pattern is represented in the second part of the table \ref{tab:TrafficProfile}.

%The traffic profile 3, is very similar to the second one, however in this the TC1 and TC2 have high utilization whereas the TC0 has the less use until then.  As profile 2 profile 3 has a better use of the network if used with BAM ATCS. In profile 3 of the table \ref{tab:TrafficProfile} it is possible to see this pattern.

%A characteristic that is repeated in all three profiles is that the network usage doesn't exceed 90\%. This limit should be emphasized since in the BAM policy implemented in the network one must avoid preemptions and devolution.
    
\begin{table}[ht]
\centering
\caption{Input Traffic Patterns}
%\vspace*{+5mm}
\label{tab:TrafficProfile}
\begin{tabular}{|c|c|c|c|c|c|c|}
\hline
 \textbf{Traffic Profile} & \textbf{1}	& \textbf{2} & \textbf{3} & \textbf{4}	& \textbf{5} & \textbf{6}  \\ \hline
TC0 & High & Medium & Low & \multicolumn{3}{|c|}{High} \\\hline
TC1 & Low & Low & Medium & \multicolumn{3}{|c|}{High}\\\hline
TC2 & Low & High & High & \multicolumn{3}{|c|}{High}\\ \hline
Link Load & \multicolumn{3}{|c|}{$<$ 90\%} & \multicolumn{3}{|c|}{$>=$ 90\%} \\ \hline
Indicated BAM & RDM/ATCS & ALL & ALL & \multicolumn{3}{|c|}{MAM} \\ \hline
\end{tabular}
\end{table}

  %\begin{figure}[ht]
  %    \centering
  %    \includegraphics[width=0.5\textwidth]{imagens/link_0-3_alloc_mod.png}
  %    \caption{deterministic traffic (3h)}
  %    \label{fig:perfil_trafego1-3}
  %\end{figure}

%The profiles of traffics 4, 5 and 6, present similar characteristics, since all of them have the characteristic of overloading the link. It can be seen in the table \ref{tab:TrafficProfile} that in the three profiles (4,5 and 6) TC0 and TC1 have a very high load. The TC2 starts with a medium/high load, reduces slightly in the second moment and returns to have a high load in period 6.

\section{Evaluated Results} \label{results}

%\begin{figure}[ht]
%      \centering
%      \includegraphics[width=0.5\textwidth]{imagens/link_3-6_alloc_mod.png}
%      \caption{deterministic traffic (6h)}
%      \label{fig:perfil_trafego3-6}
%  \end{figure}

 %\begin{figure}[ht]
 %     \centering
 %     \includegraphics[width=0.5\textwidth]{imagens/simulacao_alloc_24.png}
 %     \caption{Simulation with deterministic traffic 24h.}
 %     \label{fig:simulacao_alloc_24}
 % \end{figure}

 The proof of concept has 2 phases. In phase 1, each BAM model (MAM, RDM and ATCS - Table II) is used statically (no reconfiguration) for a 24h simulation period. Following that, BAMCBR is used to configure BAM models for the same 24h simulation period and input traffic pattern (BAMCBR - Table \ref{tab:ComparativoBAMs}). Performance metrics are captured, allowing to compare BAMs individual network performance metrics with CBRBAM cognitive management and infer about the learning process.

 In phase 2, CBRBAM pilots BAM reconfiguration process for a 24h period and the objective is to infer further details about the learning process with CBR (Table \ref{tab:AprendizadoBAMCBR}).

 BAMCBR uses the following management goals and requirements (Figure 1): i) to minimize preemption and devolution; and ii) to maximize network utilization in terms of achieving the maximum possible number of established LSPs. For all simulations using BAMCBR, MAM is always the initial BAM model configured and the "cases database" is empty (no previous manager's knowledge is inserted).

 %Das políticas apresentadas, para esta prova de conceitos determinamos como critérios de performance e otimização : (1) minimizar as preempções e devoluções; (2) contudo maximizar a utilização do enlace.
 
 %Ou seja a taxa de preempção e devolução deverá ser próxima a taxa de preempção e devolução do MAM (zero – 0), e a taxa de utilização próxima as taxas do ATCS . Sendo assim um gestor que implemente essa política espera aproveitar o melhor do MAM e do ATCS.

 Table \ref{tab:ComparativoBAMs} illustrates network performance metrics (preemption, devolution, blocking and unbroken LSP rates) for phase 1. The 3 first lines basically confirm MAM, RDM and ATCS overall expected behaviors and are relevant to check further investigated BAM cognitive management. MAM (1st line) do not share bandwidth and, as such, no preemption and devolution occurs. MAM disadvantages in relation to other models (RDM and ATCS) is that blocking rate is higher and less LSP are established (effectively the unbroken ones). RDM (2nd line) shares bandwidth and, as such, preemption do occur. Blocking rate is reduced and more LSP are effectively established in relation to MAM. ATCS (3rd line) allows generalized sharing and preemption and devolution occurs. ATCS, as expected, presents the smallest blocking rate, higher preemption and the highest number of established and unbroken LSPs.

% Na tabela \ref{tab:ComparativoBAMs}  é possível ver as estatísticas das métricas de desempenho da rede utilizando cada modelo de alocação de banda. Na primeira linha é possível perceber que não existe preempção e nem devolução, justamente por se tratar da utilização do BAM MAM. Também é possível perceber que este BAM apresenta a pior de todas as taxas de utilização da rede.
 
% A segunda linha é possível ver o desempenho da rede com a utilização do RDM e na linha três com a utilização do ATCS. Claramente é possível ver que o ATCS possui a melhor taxa de utilização da rede contudo tem os maiores índices de preempção e devolução.

%É possível observar que com a utilização do BAMCBR as exigências inicialmente apresentadas foram atendidas – baixa preempção e devolução e alta utilização.

\begin{table}[ht]
\centering
\caption{BAMs and BAMCBR Performance Metrics}
%\vspace*{+5mm}
\label{tab:ComparativoBAMs}
\begin{tabular}{|c|c|c|c|c|c|}
\hline
 \textbf{BAM} & \textbf{Preemption}	& \textbf{Devolution} & \textbf{Blocking} & \textbf{Unbroken}  \\ \hline
MAM & 0 & 0 & 24638 & 53357 \\\hline
RDM & 3813 & 0 & 15837 & 58349 \\\hline
ATCS & 3456 & 2431 & 7385 & 64721 \\ \hline
BAMCBR & 88 & 158 & 15523 & 62226 \\ \hline
%CBR2 & 32 & 99 & 15790 & 62074 \\ \hline
\end{tabular}
%\vspace{1ex}
%     \raggedright *Métricas apresentadas em total de LSPs na simulação.
\end{table}

%A cada interação, entre a rede e o BAMCBR, o modulo de aprendizagem avalia a rede e aprende com ela qual o melhor BAM para o momento com base nas políticas definidas pelo gestor.

Line BAMCBR (Table II) shows network performance with BAMCBR managing BAM configuration for the entire 24h simulation period. Result shows that preemption and devolution reduction goals were effectively achieved. In relation to the high utilization goal, BAMCBR result has a lower number of unbroken LSPs  in relation to ATCS model individually. That indicate the cognitive BAMCBR management effectively succeeded to get the best possible network utilization keeping preemption and devolution at their minimum values at the cost of having less LSPs established.

Result in Table II endorses two evidences: i) network utilization was maximized keeping preemption and devolution lower values when compared to RDM and ATCS models individually; and ii) this result can only be achieved in case BAMCBR effectively learns about network performance metrics and reconfigure BAMs to achieve manager's goals.

Phase 2 simulation results are presented in Table III and presents further evidences on how BAMCBR module learns during 4 cycles of 6 hour simulation (CBR 1/4 to CBR 4/4).

\begin{table}[ht]
\centering
\caption{BAMCBR Learning Evidences}
%\vspace*{+5mm}
\label{tab:AprendizadoBAMCBR}
\begin{tabular}{|c|c|c|c|c|c|}
\hline
 \textbf{BAM} & \textbf{Preemption}	& \textbf{Devolution} & \textbf{Blocking} & \textbf{Unbroken}  \\ \hline
BAMCBR 1/4 & 30 & 107 & 3753 & 15535 \\\hline
BAMCBR 2/4 & 62 & 36 & 3808 & 15519 \\\hline
BAMCBR 3/4 & 11 & 30 & 3888 & 15496 \\ \hline
BAMCBR 4/4 & 11 & 30 & 3888 & 15496 \\ \hline
\end{tabular}
\vspace{1ex}
%     \raggedright *Métricas apresentadas em total de LSPs na simulação.
\end{table}

The analysis of the BAMCBR 1/4 period indicates that, initially, CBR cognitive reconfiguration succeed to reduce preemption. Within BAMCBR 2/4 period, CBR cognitive management succeed to reduce devolution but at the cost of increasing preemption. At the BAMCBR 3/4 period result present an evidence that CBR cognitive management attained the best possible solution and, overall performance results keep stable for next period. This learning process also indicates that previous configuration are not used anymore by BAMCBR whenever a new configuration with better results is learned. The repetition of performance values for lines BAMCBR 3/4 and 4/4 indicates that no new case occurred and that the learning process was achieved.

\section{Final Considerations} \label{sec:final considerations}

The main evidence obtained for BAMCBR approach is that cognitive management using CBR does learn from policy-defined network current performance metrics whether the current configuration is adequate and, subsequently, is able to dynamically and autonomically reconfigure BAM models to achieve the specified manager's goal. Another relevant result obtained is that network performance was improved in alignment with manager's predefined policy.

A positive aspect of the result obtained is that the learning process occurred from the scratch. No previous knowledge was required to be inserted in the "cases database" to allow BAMCBR switch among BAM models, looking for the best configuration and find the best possible result. The drawback of not inserting any previous knowledge is that BAMCBR takes more time to find out the appropriate solution and considers eventually "bad solutions" on the learning path that are subsequently not considered (Table II learning evidences BAMCBR 1/4 to 4/4). 

Evidences obtained paved the way to autonomic network where automated decisions are necessary with the engineering of accurate knowledge from a complex network environment.

Future work will include the impact of previous knowledge utilization on the efficiency of the solution (how fast BAMCBR can reach stability), BAMCBR's dependency of the system policies defined by the manager and the evaluation of BAMCBR configuration tuning impacts for positive and negative learning issues.

\bibliographystyle{IEEEtran}

%\bibliography{ISCC2018-CBRBAM}

%\bibliography{IEEEabrv,CBRBAM}
\bibliography{CBRBAM.bib}

\end{document}